%% file: design_of_skos.tex
\documentclass[preprint]{elsarticle}
\usepackage{a4}
\usepackage{url}
\usepackage{color}
\usepackage[table]{xcolor}
\usepackage{multirow}
\usepackage{latexsym}
\usepackage{type1cm}          
\usepackage{makeidx}          
\usepackage{graphicx}         
\usepackage{multicol}         
\usepackage{framed}

\usepackage{hyperref}         

\newcommand{\urlfoot}[1]{\footnote{\url{#1}}}

\newcommand{\owl}[1]{\texttt{#1}}
\newcommand{\skoscomponent}[1]{\emph{#1}}
\newcommand{\skosrequirement}[1]{\href{http://www.w3.org/TR/skos-ucr/\#{#1}}{\sc{#1}}}
\newcommand{\skosissue}[1]{\href{http://www.w3.org/2006/07/SWD/track/issues/#1}{\sc{Issue #1}}}
\newcommand{\skosaxiom}[1]{\href{http://www.w3.org/TR/skos-reference\#{#1}}{\sc{#1}}}

\begin{document}
\title{Key Choices in the Design of Simple Knowledge Organization System (SKOS)}

\author[1]{Thomas Baker}
\author[2]{Sean Bechhofer\corref{cor1}}
\ead{sean.bechhofer@manchester.ac.uk}
\author[3,4]{Antoine Isaac}
\author[5]{Alistair Miles}
\author[3]{Guus Schreiber}
\author[6]{Ed Summers}

\cortext[cor1]{Corresponding Author}
\address[1]{Dublin Core Metadata Initiative, c/o National Library Board
  Singapore, 100 Victoria Street 14-01, Singapore 188064}
\address[2]{University of Manchester, Oxford Road, Manchester M13 9PL, UK}
\address[3]{Vrije Universiteit Amsterdam, De Boelelaan 1081a
  1081 HV Amsterdam, NL}
\address[4]{Europeana, c/o Koninklijke Bibliotheek, PO Box 90407, 2509 LK The Hague, NL}
\address[5]{Wellcome Trust Centre for Human Genetics, Roosevelt Drive, Oxford OX3 7BN, UK}
\address[6]{Library of Congress, 101 Independence Ave, SE
  Washington, DC 20540, US}

\date{\today}

\input{abstract}
\maketitle

\input{introduction}
\input{history}
\input{rationale}

\input{components}
\input{semantics}

\input{conclusion}

\input{acks}

\bibliography{bibliography}
\bibliographystyle{plainurl}

\end{document}

%% file: abstract.tex
\begin{abstract}
Simple Knowledge Organization System (SKOS) provides a data model and vocabulary for expressing Knowledge Organization Systems (KOSs) such as thesauri and classification schemes in Semantic Web applications. This paper presents the main components of SKOS and their formal expression in Web Ontology Language (OWL), providing an extensive account of the design decisions taken by the Semantic Web Deployment (SWD) Working Group of the World Wide Web Consortium (W3C), which between 2006 and 2009 brought SKOS to the status of W3C Recommendation.  The paper explains key design principles such as "minimal ontological commitment" and systematically cites the requirements and issues that influenced the design of SKOS components.

By reconstructing the discussion around alternative features and design options and presenting the rationale for design decisions, the paper aims at providing insight into how SKOS turned out as it did, and why. Assuming that SKOS, like any other successful technology, may eventually be subject to revision and improvement, the critical account offered here may help future editors approach such a task with deeper understanding.

\end{abstract}

%% file: introduction.tex
\section*{Introduction}
\label{sec:intro}

Simple Knowledge Organization System (SKOS) --- a vocabulary and data model for
expressing Knowledge Organization Systems (KOSs) such as thesauri and
classification schemes for referencing and re-use in Semantic Web applications
--- was developed by successive projects and working groups from the late 1990s
through its publication in August 2009 as a World Wide Web Consortium (W3C)
Recommendation.\urlfoot{http://www.w3.org/2009/07/skos-pr}  This paper
describes the work of the W3C Semantic Web Deployment Working Group, which was
chartered in 2006 to carry SKOS Core, a W3C Working Draft, through the rigorous
review required by the W3C Recommendation Track
process.\urlfoot{http://www.w3.org/2006/07/swdwg-charter}

The final results of that process are recorded in the formal specification for
SKOS~\cite{skos}. This paper, in contrast, focuses on the process itself.  By
reconstructing the discussion around alternative features and design options
and presenting the rationale for key decisions, the paper aims at providing
insight into how SKOS turned out as it did, and why.  Assuming that SKOS, like
any other successful technology, may eventually be subject to revision and
improvement, the critical account offered here may help future editors approach
such a task with deeper understanding.

After presenting a brief history of SKOS from 1997 through 2009, the paper
outlines the rationale for a language, other than existing formal ontology
languages, for expressing Knowledge Organization Systems.  Drawing a contrast
between logically precise conceptual structures and more intuitive, pragmatic
knowledge representations, the section describes the principle of ``minimal
ontological commitment'' that guided the design of SKOS.  

The middle section of the paper walks through the components of the SKOS model
--- SKOS \skoscomponent{Concepts} (and how they differ from formal-ontological
Classes), \skoscomponent{Concept Schemes}, \skoscomponent{Semantic Relations}
between concepts, \skoscomponent{Lexical Labels}, \skoscomponent{Documentation
Properties}, and \skoscomponent{Collections} of concepts.  The section
considers several proposed features of SKOS deemed by the working group to be
out of scope.

While SKOS was developed for expressing KOSs as ``concept schemes'' --- sets of
interrelated concepts --- without modeling those concepts as formal
``classes,'' the data model for SKOS itself is defined as an ontology, i.e., as
a set of formal properties and classes expressed using the W3C Web Ontology
Language (OWL).  The final section of the paper reviews the semantics of SKOS
properties and classes as defined by axioms supporting inference and
``integrity conditions'' for when data can be considered ``not consistent''
with the SKOS data model.  The section also considers the compatibility of SKOS
with different variants of OWL and with previous versions of SKOS itself.  The
paper concludes by highlighting issues that could provide starting points for a
future revision of the specification.


%% file: history.tex
\section{History of SKOS from the late 1990s through 2009}
\label{sec:history}

Today's SKOS can be traced back to work on improving search interfaces in the
European project Desire (1997--2000).  The original W3C Resource Description
Framework (RDF) Model and Syntax Working Group (1997--1999), aware of Desire,
raised the question of expressing thesauri in RDF as an
issue.\urlfoot{https://www.w3.org/RDF/Group/Schema/openissues.html}
Phil Cross, Dan Brickley, and Traugott Koch turned the Desire results into a
proposal, published jointly by the Institute for Learning and Research
Technology (ILRT) in the UK and the Lund University Library Netlab in Sweden,
``for encoding a core set of thesaurus relationships using an RDF
schema.''\urlfoot{http://www.ilrt.bristol.ac.uk/publications/researchreport/rr1011/report_html?ilrtyear=00}
This draft schema was picked up by the European project LIMBER (Language
Independent Metadata Browsing of European Resources, 1999--2001), which defined
a vocabulary based more explicitly on ``concepts'' labeled by terms in multiple
languages.\urlfoot{http://journals.tdl.org/jodi/article/view/32/33}

The results of the LIMBER Project fed into the SWAD Europe project (Semantic
Web Advanced Development,
2001--2004).\urlfoot{http://www.w3.org/2001/sw/Europe/} In SWAD Europe,
Alistair Miles of Rutherford Labs solicited input from experts on thesaurus and
classification standards, creating a community of interested users, for whom W3C set up a
community mailing list,
public-esw-thes,\urlfoot{http://lists.w3.org/Archives/Public/public-esw-thes/}
and the revised vocabulary was published under the name ``Simple Knowledge
Organization System.''  This draft was picked up in 2004 by the W3C Semantic
Web Best Practice and Deployment Working Group (2004--2006), whose Porting
Thesauri Task Force\urlfoot{http://www.w3.org/2004/03/thes-tf/mission}
created a home page for what was now called ``SKOS Core.'' 
In 2005, the working
group published ``SKOS Core Vocabulary Specification'' as a W3C Working
Draft.\urlfoot{http://www.w3.org/TR/2005/WD-swbp-skos-core-spec-20051102/}
SKOS Core was taken as a starting point for the review process described in
this paper.

The Semantic Web Deployment Working Group began by distilling requirements for
SKOS out of use cases solicited from early adopters about present and future
applications~\cite{skosuse}.  Successive revisions of the 2005 SKOS Core
specification were posted for public comment as Working Drafts, then as Candidate
and Proposed Recommendations, prior to finalization as a W3C
Recommendation.  The editors of the specification were supported by two
working-group chairs with active input from a dozen or two working-group
members and a wider circle of external reviewers and mailing-list followers.
Discussion took place on the working group's mailing
list\urlfoot{http://lists.w3.org/Archives/Public/public-swd-wg/} and on
public-esw-thes for the wider
community.\urlfoot{http://lists.w3.org/Archives/Public/public-esw-thes/}
The group met over a period of 35 months in three face-to-face meetings and 110
near-weekly teleconferences.  Teleconferences used W3C's bot-supported
telephone bridge, which assigned URIs to actions ``scribed'' into a shared chat
channel and automatically generated draft minutes, complete with pointers to
the agenda, previous minutes, actions past and current, mailing-list postings,
and document drafts.  As technical or design issues were formally raised they
were assigned URIs and added to an Issue Tracker
\urlfoot{http://www.w3.org/2006/07/SWD/track/issues/} that automatically
collected links to any minutes or postings in which the issues were mentioned.
Each such URL cited in this paper leads the interested reader into a web of
richly interlinked working-group resources.

The following discussion will make reference to online resources
produced during the WG process. Rather than peppering the narrative
text with URIs, references to issues, requirements and axioms will be
handled as follows:

\begin{itemize}

\item{\textbf{Issues}}. Details of all issues are documented in the Working
Group's issue tracker at \url{http://www.w3.org/2006/07/SWD/track/issues}.
Issues will be cited in the text by number, e.g., \skosissue{27}.

\item{\textbf{Requirements}}. Requirements are documented in the SKOS Use Cases
and Requirement document at \url{http://www.w3.org/TR/skos-ucr}. Requirements
will be referred to in the text by their handles, 
e.g., \skosrequirement{R-GroupingInConceptHierarchies}.

\item{\textbf{Axioms}}. SKOS axioms are listed in Tables~\ref{tab:intaxioms}
and~\ref{tab:xlaxioms}. Details of these axioms are given in the SKOS Reference
document at \url{http://www.w3.org/TR/skos-reference}. Axioms will be referred
to in the text by their ``S'' handle, e.g., \skosaxiom{S1}.

\end{itemize}

For all of  the above, full URI references will be available in digital versions of the
paper.


%% file: rationale.tex
\section{Rationale for SKOS}
\label{sec:rationale}

Many institutions develop and maintain Knowledge Organization Systems (KOSs)
--- thesauri, classification systems, subject heading lists, folksonomies, and
the like, holding concepts and terminologies for a wide range of domains --- as
backbone structures for their information systems.  The potential of such
KOSs to serve as components in knowledge-rich applications has been
recognized since the rise of the Web in the 1990s.

Porting an existing KOS for use in Semantic Web applications, however, is not a
trivial problem. The Semantic Web languages for expressing domain knowledge are
mathematically formal in nature.  The vocabulary description language of the
Resource Description Framework (RDFS) and the Web Ontology Language (OWL), in
particular, provide ways to define classes and properties and to associate
those classes and properties with formal reasoning rules that enforce
constraints or produce new knowledge by inference. KOSs, on the other hand,
have typically been designed not as formally precise representations of domain
knowledge, but as informal structures reflecting the intuitive knowledge of
human users in a form useful for resource discovery (e.g., through supporting
query expansion).  KOSs have variously been classified as ``term-based'' or
``concept-based'' depending on how explicitly they are intended to represent
conceptual structures.\footnote{The evolution of standards such as
ISO2788~\cite{iso2788} (into ISO25964~\cite{iso25964}) illustrates the shift,
but also the continuity from one representation approach to the other.}
Traditional KOS standards have never included the sort of formal axioms
expressed by Semantic Web ontology languages.  

Informally defined KOSs cannot typically be translated into the language of
RDFS and OWL properties and classes, with their formal-logical implications,
without introducing potentially false or misleading logical precision.
Informal KOSs may be converted into formal ontologies (see~\cite{hyvonen}),
but the process of assigning appropriate formal semantics to the elements of a
KOS may require a long, hard modeling effort.  Hierarchical relationships, for
example, must be disambiguated into relationships of class instantiation, class
subsumption, part-whole, or other types --- a process that cannot usually be
automated.  An analysis of the thesaurus of the National Cancer
Institute~\cite{nci} (as reported in~\cite{ceuster}), for example, found
conceptual structures that are incompatible with formalized frameworks that
assume stricter modeling principles.  The AGROVOC thesaurus of multilingual
agricultural terminology, the product of many people over many years working
from multiple perspectives, was straightforwardly converted into a hierarchy of
OWL classes many years before the finalization of SKOS.  While the maintainers
of AGROVOC-in-OWL intended to increase its ontological precision over time,
through editorial correction and refinement, it eventually proved to be more
practical simply to convert AGROVOC back into the formally less ``committed''
form of a SKOS concept scheme, leaving it to designers of specific
implementations to upgrade parts of the thesauri into class-based ontologies
when required to support reasoning~\cite{baker-agrovoc}.

The traditional use cases for which KOSs were typically designed are still
relevant in the Web context.  One key role of a controlled
vocabulary, for example, is to improve precision when retrieving objects from
an indexed collection.  The hierarchical and associative relationships of
thesauri enable users to browse for search terms, and information retrieval
applications can use this structure to automatically expand queries, which improves
recall.  Applications such as simple search or browsing of documents or
``conceptual spaces'' can all benefit from a shared basis for data
exchange and linking.  For such purposes, Semantic Web technology is indeed a game
changer, as it allows users and developers to seamlessly re-use data from
different contexts, or to link together multiple KOSs, in order to achieve
broader or deeper search, even across languages. Expressing KOSs as Linked Data
allows the library community to create pools of trusted URIs citable by
catalogers in resource descriptions in support of such
applications~\cite{lld-report,lld-usecases}.


SKOS aims at providing a path for migrating KOSs to a Semantic Web context at
low cost by expressing features common to a wide range of KOS types.  The SKOS
properties for ``broader,'' ``narrower,'' and ``related,'' for example, are intended to
capture the native, sometimes ambiguous semantics of existing thesauri and
similar structured vocabularies. Using SKOS, no additional intellectual work is
required to represent these relationships in RDF, allowing the maintainers of
controlled structured vocabularies to leverage their existing investments.

The design of SKOS followed the principle of
making a \emph{minimal ontological commitment} to the nature of concepts and of
relationships between concepts.  As explained by Thomas Gruber~\cite{gruber}:

\begin{quote}
An ontology should require the minimal ontological commitment sufficient to
support the intended knowledge sharing activities. An ontology should make as
few claims as possible about the world being modeled, allowing the parties
committed to the ontology freedom to specialize and instantiate the ontology as
needed.\end{quote}

The principle of avoiding over-commitment guided many of the discussions about
possible extensions to SKOS.  Where the use cases collected by the working
group demonstrated no clear requirement for a candidate feature, or in the
absence of clear usage experience, the group tended to opt for a ``safe''
course of action.  As a result, SKOS captures the basic, informal semantics
most commonly required by the use cases.  Where there was doubt that a
particular feature would be easy to understand or use, the working group
generally chose to omit the feature from the specification.

The working group was particularly focused on keeping SKOS compatible with the
thesaurus standards ISO 2788 and ISO 5964,\footnote{The SKOS Primer includes a
table of correspondences with ISO 2788 and ISO 5964
\url{http://www.w3.org/TR/skos-primer/\#seccorrespondencesISO}.} with the result that the SKOS data
model reflects standard thesaurus construction principles.  SKOS does not,
however, express all of the best practices described in the ISO standards, nor
does it include the elements needed to capture all of the features of any
given, existing KOS standard, such as specializations of broader and narrower
hierarchical relations (see Section ~\ref{sec:specializations-hierarchical-relations}).  Experience indeed
shows that best practices are not always followed---a problem revealed, for example, when generic
``See also'' references in the Library of Congress Subject Headings were converted
into standardized thesaurus relations~\cite{spero}---and that some
KOSs use idiosyncratic constructs for meeting very specific requirements. 
The working group felt that
fully committing SKOS to supporting the creation and validation of any
particular type of concept scheme, such as a standard thesaurus, would create
an obstacle to the wide-spread adoption of SKOS by users of other types.  

Lightly specified by design, SKOS is intended to prevent data publishers from
introducing false precision into their data and to prevent inference engines
from drawing unwarranted conclusions.   In some cases, however, the
specification recommends usage conventions, such as best practices for KOS
design.  The SKOS model thus presents two layers of specification: formal,
enforceable axioms, along with weaker ``guidelines.''  Guidelines are not
represented formally, nor they are considered to be inviolable integrity
constraints; rather, they are considered to be advisory. 

Opting for such a minimal approach is made dramatically easier by the
vocabulary extension mechanisms offered natively by Semantic Web technology.
Applications that require more constrained behaviour may define compatible
extensions to SKOS~\cite{skosprimer}. For example, modelers may coin
sub-classes and sub-properties of SKOS properties or associate those
properties with specific formal axioms.  The RDF data model allows
properties from such extension vocabularies to be used alongside properties
from SKOS in expressing data.  Where properties seen as required were already
provided elsewhere, such as the Dublin Core property \owl{dc:subject}, the
working group deferred to existing vocabularies.

%% file: components.tex
\section{Components of SKOS}
\label{sec:2}


\begin{framed}
Using SKOS, 
\textbf{concepts} can be identified using URIs, 
\textbf{labeled} with lexical strings in one or more natural languages, 
assigned \textbf{notations} (lexical codes),
\textbf{documented} with various types of note, 
\textbf{linked to other concepts} and organized into informal hierarchies and association networks,
aggregated into \textbf{concept schemes}, 
grouped into labeled and/or ordered \textbf{collections}, and 
\textbf{mapped} to concepts in other schemes.
\end{framed}

\begin{figure}
  \centering
    \includegraphics[width=\textwidth]{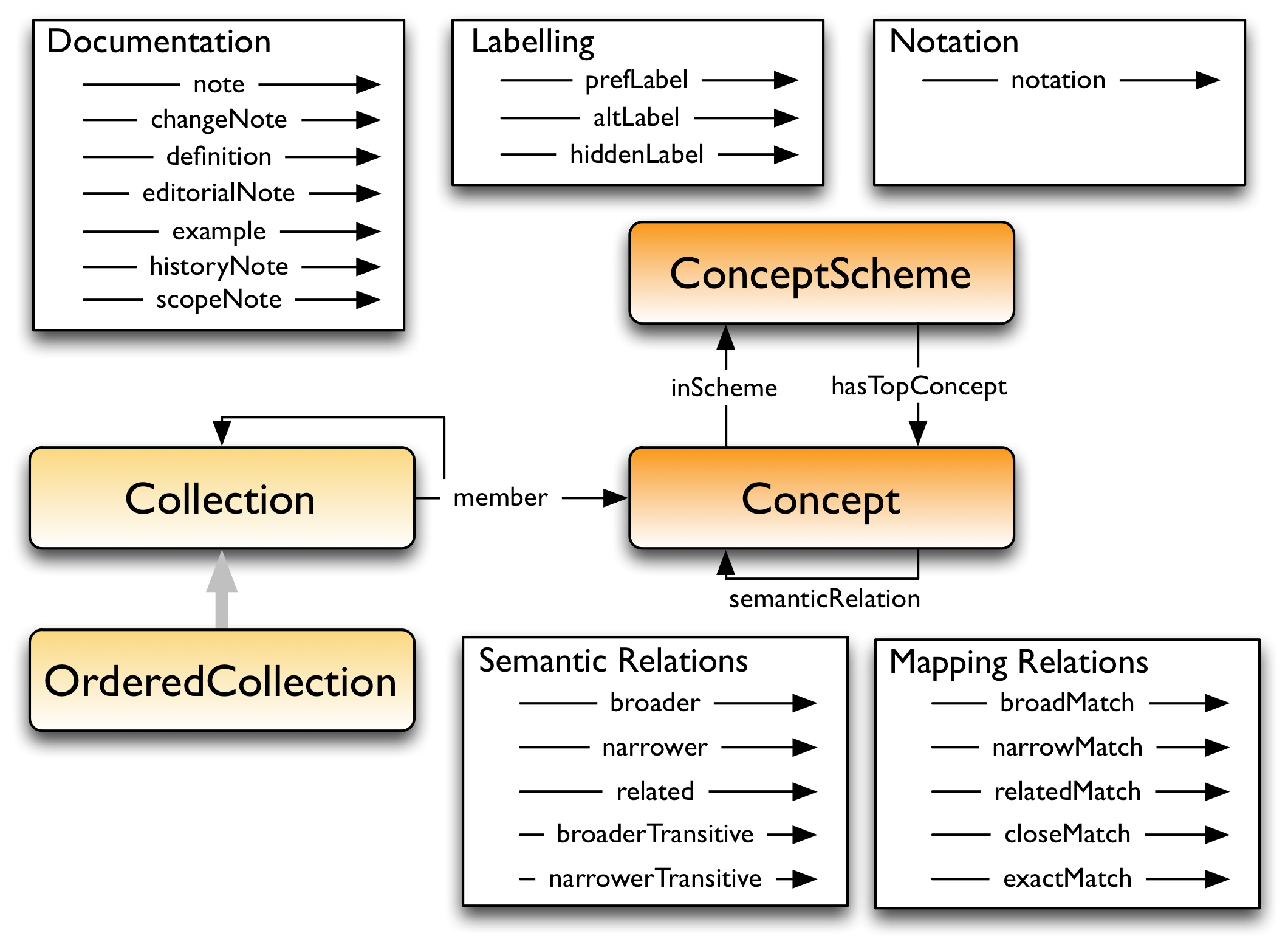}
      \caption{Main elements of the SKOS data model}
  \label{fig:skos-model}
\end{figure}

The SKOS data model enables features listed above---identifying, labeling,
documenting, linking, and mapping concepts, and aggregating concepts into
concept schemes or collections---by defining the elements depicted in Figure
~\ref{fig:skos-model}. This section looks at the design choices made in
modeling those components.


\subsection{SKOS Concepts (and how they differ from OWL Classes)}
\label{sec:concepts}

\paragraph{A wide diversity of concepts.}

SKOS is designed to express, in an interoperable way, different types of
Knowledge Organization System---sets of terms or concepts, whether listed
with definitions (glossaries), in hierarchical structures (basic
classifications or taxonomies), or characterized by more complex semantic
relations (thesauri, subject heading lists, or other advanced structures). Each
type of KOS has its own specific characteristics. Yet they all organize
knowledge by gathering a coherent set of lexical entities (terms,
words, headings, captions\ldots) around more abstract
notions that the SKOS model represents as \skoscomponent{Concepts}. In a thesaurus, for
example, a concept is the construct that clusters a \emph{preferred term} (the
one used for describing resources in a document retrieval system) with
near-synonymous \emph{alternative terms} (or variants).  A KOS may link
such concepts among themselves with various types of semantic relations, such
as class-subclass, part-whole, or looser associative links.

SKOS leaves ample room for interpreting the notion of concept, and many
artifacts from information science and other fields fall in scope. As the SKOS
Reference puts it~\cite{skos}, ``a SKOS concept can be viewed as an idea or
notion; a unit of thought. However, what constitutes a unit of thought is
subjective, and this definition is meant to be suggestive, rather than
restrictive.''  

If the objective is information retrieval via the use of a knowledge
organisation system as a subject indexing language, then one can take an
operational view and define concepts as units of indexing and
retrieval~\cite{svenonius2000:foundation}.  The subject indexing process
can then be viewed as the action of linking documents (such as a textbook
about butterflies) to concepts (such as a concept labeled ``butterflies''), and
the retrieval process involves selecting one or more concepts to use as a
subject query and retrieving the sets of documents linked to those
concepts. One possible formulation of this view is to describe the set of
documents linked to the same concept as a kind of ``document extension'' of
that concept.  Some approaches to mapping are based on this notion, as when
conceptual equivalences between concepts are derived by measuring the
overlap between the document extensions with which they are
associated~\cite{stitch-extensionalalignment}.


Information retrieval use cases are an important motivation for SKOS, and
the SKOS data model is perfectly compatible with this view. However,
subject indexing and retrieval are not the only uses for knowledge
organisation systems, so SKOS does not attempt to normatively
define or formalize any relationship between documents and concepts. This
flexibility enables SKOS to represent knowledge organisation systems used
in a variety of applications, as well as enabling implementers of
information retrieval systems to explore alternative retrieval strategies
and algorithms such as query expansion.
  
Moreover, conceptual vocabularies need not be intended primarily for describing
documents for information retrieval.  
At the most basic level, applications merely require that a concept have
identity and that it have features which distinguish it from other concepts
within a KOS, such as natural-language labels, definitions, and semantic
relations to other concepts.  The group felt that this simple, flexible
model would cover most of the available use cases and requirements while
enabling a broad range of applications, whereas formally reconciling
traditional KOS models at a higher degree of granularity would have been
both more difficult and of less obvious utility. 
     
\paragraph{SKOS concepts vs. OWL classes.} The concepts from Knowledge
Organization Systems (and hence, SKOS concepts) are often wrongly interpreted
as classes from formal ontology languages like OWL.  Some SKOS concepts indeed
reflect universal categories that also appear in in OWL ontologies, such as
``animals'' and ``cats'' in a vocabulary animals.  Yet, as seen previously,
SKOS concepts are not by default provided with precise extensional semantics,
while an OWL class explicitly describes a collection of individuals.  Following
the example above, the interpretation of an OWL Class ``Butterfly'' would be
the set of butterflies. 

From a technical perspective, SKOS \skoscomponent{Concepts} are simply
individuals in an OWL ontology (see Section~\ref{sec:sem}).  These
individuals are interpreted as arbitrary elements in the domain that
\emph{might} (or might not) correspond to collections of indexed documents.


Hierarchical relationships between SKOS \skoscomponent{Concepts},
moreover, \emph{might} (or might not) correspond to sub-class relationships
between OWL classes (\owl{owl:Class}).  Where a KOS might assert
\skoscomponent{broader} links between ``Dog,'' ``Collie'' (a type of dog),
and ``Lassie'' (a particular dog), an OWL ontology might specify that
``Collie'' is a sub-class of ``Dog'' (\owl{rdfs:subClassOf}), then
``Lassie'' could be described as an instance of ``Collie''
(\owl{rdf:type}).  Some KOS standards distinguish a class-instance variant
of ``broader,'' such as the ``broader instantive'' relationship of ISO
2788~\cite{iso2788}.  (Indeed, a preliminary draft SKOS Extensions
Vocabulary Specification with ``broader instantive'' and other such
variants of the semantic relation properties was posted for discussion in
2004.\urlfoot{http://www.w3.org/2004/02/skos/extensions/spec/2004-10-18.html})
Many KOSs, however, use the same general relation for all hierarchical
links.

As the SKOS Reference says, concept data could be ``facts about the thesaurus
or classification scheme itself, such as `concept X has preferred label ``Y'' '; 
not facts about the way the world is arranged within a particular subject
domain, as might be expressed in a formal ontology.'' 
As hinted in Section~\ref{sec:rationale}, converting a traditional KOS into
an OWL ontology may thus require some hard ``cleaning.''  Conversion into a
straightforward, lightweight representation is often the most appropriate
course of action~\cite{baker-agrovoc,vanAssem-thesis}.  One may see formal (OWL)
ontologies as KOSs, as they organize knowledge in a domain; but clearly not all
KOSs can be directly interpreted as formal ontologies.

\paragraph{Disjointness of the SKOS Concept Class.}
The previous
paragraph highlights that KOS concepts function in a quite specific way: one
can view them as proxies that establish an ``indirection layer'' between
lexical entities and ``real-world'' ones, either represented as individuals
(say, a person) or classes (say, all persons with a specific role) in the OWL
sense.  A test for identifying such resources could be for example the ``date
of creation'' associated with them. The resource that stands for a person in a
name authority file (thus, represented as an instance of SKOS concept) will
probably have a different date associated to it than the one associated to the
resource that stands for that person as a ``real person'' (represented using
the \owl{foaf:Person} class).\footnote{For example, the data available for
an authority name for Michelle Obama indicate a creation date of 2008, quite
some time after Michelle Obama was born. See
\url{http://id.loc.gov/authorities/names/n2008054754} and a discussion on the
SKOS community mailing list,
\url{http://lists.w3.org/Archives/Public/public-esw-thes/2009Nov/0000.html}}

This observation leads to a first kind of representation pattern, which
distinguishes separate KOS concepts from entities in the real world (or classes
thereof) and represent the referential link between them using properties such
as \owl{foaf:focus} (defined as ``The underlying or `focal' entity
associated with some SKOS-described
concept.''\urlfoot{http://xmlns.com/foaf/spec/\#term_focus}). This is
for example what was retained for the Virtual International Authority
File (VIAF),\urlfoot{http://viaf.org/} which creates for each cluster of
authority records an instance of \owl{foaf:Person} and (at least) one
instance of \owl{skos:Concept}, linking the latter (e.g., \url{http://viaf.org/viaf/sourceID/SELIBR\%7C317488#skos:Concept}) to the former (\url{http://viaf.org/viaf/85312226}) using \owl{foaf:focus}.

Such approaches would fit well a modeling choice making
\owl{skos:Concept} disjoint with other classes of entities, such as
\owl{foaf:Person} or meta-modeling classes like \owl{owl:Class}. This
would rule out that different ``modelling streams,'' each coming with
different kind of possibly incompatible data, are ``crossed'' within
one same graph.  (The cultural reference in working-group discussions
on this topic was that of the dire warning, from the 1984 film
\emph{Ghostbusters},\urlfoot{imdb:ghostbusters} never to ``cross the
streams'' of proton beams from multiple particle throwers because,
vaguely but ominously, ``it would be bad.'')

However, the working group opted for not asserting explicit disjointness
between SKOS concepts and non-SKOS classes. The first reason is quite
pragmatic: the world of ontologies beyond SKOS is wide, and choosing specific
classes for disjointness statements would have been an incomplete, biased
effort. \owl{skos:Concept} is only formally disjoint with other
classes in the
SKOS namespace: \owl{skos:Collection}, \owl{skos:ConceptScheme},
\owl{skosxl:Label}, which will be introduced below. Advocates of the
disjointness option may argue that SKOS could have featured a new class of
``Non-Concepts'' to handle the case, next to a property to relate the two
disjoint categories.\footnote{The SKOS Primer suggests to use a dedicated
(annotation) property like \owl{ex:correspondingConcept}. A \owl{skos:it}
was also proposed.  It was only after SKOS was published as a recommendation
that \owl{foaf:focus} emerged as possible standard candidate.} At that time,
however, the group felt that the proposed patterns were still not mature enough
and quite out of the scope defined in our charter, because such information is
usually not present in the data defining concepts in existing KOSs.

The second reason for not declaring SKOS concepts to be disjoing with
``non-concepts'' stems, again, from the requirement for minimal commitment, as
well as from a concern not to rule out valid patterns. As pointed out above,
a \owl{skos:Concept} is intended to provide a neutral target for migrating a wide
diversity of KOS concepts to the Web of Data. This includes cases where
concepts are also elements of formalized ontologies. Some OWL properties and classes
can be seen as members of a concept scheme as, for example, in applications that would not
handle the full complexity of OWL reasoning but would require lexical
annotations richer than those supported by OWL. The Library of Congress, for example,
represents MARC relators\urlfoot{http://id.loc.gov/vocabulary/relators}
both as SKOS concepts and OWL properties. The SKOS Primer discusses cases in which
it might make sense to treat an instance of SKOS \skoscomponent{Concept} also as a 
class.\urlfoot{http://www.w3.org/TR/skos-primer/\#secskosowl}

\subsection{Concept Schemes} \label{sec:schemes}

\begin{framed}
Using SKOS, 
\textbf{concepts} can be {\color{gray} identified using URIs, 
\textbf{labeled} with lexical strings in one or more natural languages, 
assigned \textbf{notations} (lexical codes),
\textbf{documented} with various types of note, 
\textbf{linked to other concepts} and organized into informal hierarchies and association networks,
}
aggregated into \textbf{concept schemes}, 
{\color{gray}
grouped into labeled and/or ordered \textbf{collections}, and 
\textbf{mapped} to concepts in other schemes.
}
\end{framed}

Sets of concepts were referred to as ``schemes'' as early as the Limber Project
(1999--2001).\urlfoot{http://www.w3.org/2001/sw/Europe/reports/thes/8.2/}  It
is worth noting that the Dublin Core community began talking in 1997 about
``schemes,''\urlfoot{http://www.dlib.org/dlib/june97/metadata/06weibel.html}
one type of which came to be called a ``vocabulary encoding
scheme''\urlfoot{http://dublincore.org/usage/documents/2003/02/07/principles/}---a
notion much less specific than, but not incompatible with, the SWAD Europe
project's notion of a SKOS \skoscomponent{Concept Scheme}.

The ``ability to explicitly represent the containment of any SKOS individual or
statement within a concept scheme'' was accepted by the working group as a
candidate requirement for SKOS (\skosrequirement{R-ConceptSchemeContainment}).
The ability to express the containment not only of particular concepts, but
also of particular statements using SKOS predicates, such as
\owl{skos:broader}, was seen as necessary for tracking the provenance of a
concept scheme's informational content, for example to establish trust.  Two
properties from the 2005 SKOS Core specification---\owl{skos:inScheme} and
\owl{skos:hasTopConcept}---already provided a way to relate SKOS concepts
(along with instances of other classes, such as \owl{skos:Collection}) to a
given concept scheme.  

Expressing the containment of statements, on the other hand, implied a
mechanism for denoting an entire set of statements as a named entity---a
challenge faced by any RDF-based application and thus not specific to SKOS.
The use cases for containing relations between concepts also seemed more
marginal than for containing concepts.  The text used to close this issue
(\skosissue{36})and Section 5.3 of the SKOS Primer point to ongoing work on
named graphs and RDF Datasets in
SPARQL\urlfoot{http://http://lists.w3.org/Archives/Public/public-swd-wg/2008Jan/0093.html}---work
which at the time of writing in 2012 remains on the agenda of the W3C working
group developing RDF 1.1.\urlfoot{http://www.w3.org/2011/rdf-wg/}

\subsection{Semantic Relations} \label{sec:relations}

\begin{framed}
Using SKOS, 
\textbf{concepts} can be {\color{gray} identified using URIs, 
\textbf{labeled} with lexical strings in one or more natural languages, 
assigned \textbf{notations} (lexical codes),
\textbf{documented} with various types of note, 
}
\textbf{linked to other concepts} and organized into informal hierarchies and association networks,
{\color{gray}
aggregated into \textbf{concept schemes}, 
grouped into labeled and/or ordered \textbf{collections}, and 
\textbf{mapped} to concepts in other schemes.
}
\end{framed}

The properties \owl{skos:broader}, \owl{skos:narrower} and
\owl{skos:related} are referred to collectively as the SKOS semantic relation
properties. They have their origins in thesauri---
controlled structured vocabularies used primarily for keyword indexing of
collections of documents or other objects~\cite{iso2788}. Thesaurus standards provide
guidance on the use of hierarchical and associative relationships when
constructing a thesaurus.  However, because these relationships exist primarily
as aids to information retrieval, some ambiguity is permitted. There has been
no need for them to support precise formal entailments (such as those
supported, for example, by a class subsumption hierarchy in an ontology).
Given this context, a number of design decisions needed to be made
during the standardisation of SKOS concerning the formal definition of
the properties \owl{skos:broader}, \owl{skos:narrower} and \owl{skos:related}.


Some constraints were deemed uncontroversial. For example, \owl{skos:broader} and
\owl{skos:narrower} form an inverse property pair (\skosaxiom{S25}), describing the two
directions of a hierarchical relationship.  If the concept ``mammals'' is
linked to the broader concept ``animals,'' then the concept ``animals'' is
linked to the narrower concept ``mammals.'' The property \owl{skos:related} is symmetric (\skosaxiom{S23}) because the fact that
two concepts are associated with each other is independent of direction
(although sub-properties of \owl{skos:related} may be defined as directional, i.e.,
non-symmetric).  If the concept ``birds'' is related to the concept ``ornithology,''
then the concept ``ornithology'' is related to the concept ``birds.''
In SKOS, hierarchical and associative relationships are declared to be disjoint (\skosaxiom{S27}).

Other constraints, less obvious, concerned transitivity,
sub-properties of semantic relations, reflexivity and cycles. 


\paragraph{Hierarchies and transitivity.}
\label{sec:transitivity}
It was decided that the properties \owl{skos:broader} and \owl{skos:narrower}
would not be transitive, and that by convention
these properties should only be used to assert direct (i.e.,
immediate) links between concepts. This decision was made to simplify
implementation. For example, many applications will render hierarchical
relationships as a tree, and so need some convenient way to
differentiate immediate links (parent/child) from indirect links
(ancestor/descendant). To support the fairly common use case where applications do want to use
the transitive closure of \owl{skos:broader} or \owl{skos:narrower} (e.g., to
expand a search query), transitive super-properties
\owl{skos:broaderTransitive} and \owl{skos:narrowerTransitive} were defined.
Note that exactly how retrieval applications make use of
\owl{skos:broader}, \owl{skos:narrower} and \owl{skos:related} to improve recall is not
defined by the SKOS specifications and is left to the application.
Some applications will take the view that if document D is indexed
with concept A, then document D will always be relevant to a query for
any concept that is an ancestor of concept A in the hierarchy. This is
equivalent to the view that if
a document is about growing vegetables, then the document is
necessarily also about gardening. If the property \owl{dc:subject} is taken
to represent the ``aboutness'' relationship between document and
concept, then this behaviour may easily be implemented, for example,
by computing the transitive closure of \owl{skos:broaderTransitive}.
However, other applications may take the less categorical view that relevance
is likely to degrade as a query is expanded away from some focal concept, and
that the different properties \owl{skos:narrower}, \owl{skos:broader}, and
\owl{skos:related} might correspond to different shapes or rates of
degradation~\cite{tudhope}.

\paragraph{Sub-properties of \owl{skos:broader} and \owl{skos:narrower}.}
\label{sec:specializations-hierarchical-relations}
Some thesauri disambiguate the hierarchical relationship into one of
class subsumption, instantiation, or part-whole relationship. The
working group discussed whether the SKOS standard should define
sub-properties of \owl{skos:broader}, such as the putative properties \owl{broaderGeneric} (for class
subsumption), \owl{broaderInstantive} (for class instantiation) and
\owl{broaderPartitive} (for part-whole relations), corresponding to distinctions
made in thesaurus standards~\cite{iso2788,iso25964} (see also \skosissue{56},
\skosissue{150}, \skosissue{178}).
There would be obvious value in having a standard set of properties,
rather than leaving it to third parties to define their own extensions
to the detriment of interoperability. However, the working group
was also conscious that there is overlap here with RDFS and OWL and
was reluctant to define new properties that might be redundant with
existing standards. For instance, one might have seen \owl{broaderGeneric}
as equivalent to \owl{rdfs:subClassOf} and \owl{broaderInstantive} 
to \owl{rdf:type}.\urlfoot{http://lists.w3.org/Archives/Public/public-swd-wg/2008Mar/0037.html} 
Whether or not it would be appropriate to
use such RDFS or OWL properties in these cases would require a deeper understanding of use
cases in which SKOS and OWL are used in combination. The working group
decided to postpone this decision, leaving it open for a future
revision of the standard.

\paragraph{Reflexivity and cycles.}
For a conventional thesaurus or similar vocabulary, it is an error for
a concept to be in a hierarchical relationship with itself, or to be
associated with itself (reflexivity).  It could be argued that these constraints
should become part of the SKOS data model by stating formally that
\owl{skos:broader} and \owl{skos:related} are irreflexive properties, and that
this would promote consistent implementation. Similarly, in a
thesaurus it is an error for there to exist any cycles within the
concept hierarchy, which could be enforced by declaring
\owl{skos:broaderTransitive} to be irreflexive.
However, the working group was also conscious that there are possible
advanced usage patterns (or extensions to SKOS) where both SKOS and OWL would be used together
within the description of the same knowledge organisation system, and
that more work was required to understand these patterns.
Although this may be an uncommon edge case, under some of these
patterns, it is conceivable that inferences such as ``\textless A\textgreater  \owl{skos:broader}
\textless A\textgreater'' could arise (for example, if someone were to assert that
\owl{rdfs:subClassOf} is a sub-property of \owl{skos:broader}).
It was therefore decided that no formal statements on the reflexivity
of the SKOS semantic relation properties would be made, although some
informal guidance would be given to application developers on how to
detect structural features that are likely to represent errors in the
majority of use cases.

\subsection{Mapping Relations}

\begin{framed}
Using SKOS, 
\textbf{concepts} can be {\color{gray} identified using URIs, 
\textbf{labeled} with lexical strings in one or more natural languages, 
assigned \textbf{notations} (lexical codes),
\textbf{documented} with various types of note, 
\textbf{linked to other concepts} and organized into informal hierarchies and association networks,
aggregated into \textbf{concept schemes}, 
grouped into labeled and/or ordered \textbf{collections}, and 
}
\textbf{mapped} to concepts in other schemes.
\end{framed}

The use cases for SKOS confirmed a strong requirement for mapping between
related concepts in different concept schemes.  Indeed, the prospect of
enabling machine-readable mappings between concept schemes developed in a
diversity of contexts, and possibly on the basis of different modeling
principles, was expected to be a key advantage of expressing those systems in
the common language of SKOS.  Taking as its starting point an unfinished 
SKOS Mapping Vocabulary
Specification from 2004,\urlfoot{http://www.w3.org/2004/02/skos/mapping/spec/2004-11-11.html}
the working group settled on five mapping properties: \owl{skos:broadMatch},
\owl{skos:narrowMatch}, \owl{skos:relatedMatch}, \owl{skos:closeMatch}, and \owl{skos:exactMatch}, all
of which were declared, either directly or by inference, to be sub-properties
of \owl{skos:mappingRelation}, itself a sub-property of 
\owl{skos:semanticRelation}.  

Much of the discussion about mapping properties revolved around clarifying how
they differed from analogous semantic relation properties.  The mapping
properties \owl{skos:broadMatch}, \owl{skos:narrowMatch}, and \owl{skos:relatedMatch} were
declared to be sub-properties, respectively, of \owl{skos:broader}, \owl{skos:narrower},
and \owl{skos:related}.  However, these ``parallel'' properties were not otherwise
distinguished in a formal sense.  The question indeed arose whether, given this
lack of formal-semantic distinction, separate properties for broader, narrower,
and related matches were needed at all.

The dilemma, as the working group saw it, was that large parts of the KOS
community saw inter-KOS mapping relations and intra-KOS semantic relations as
fundamentally different things, perhaps even disjoint from each other.  From
the standpoint of the working group, the intended distinction between
mapping relations and semantic relations depended, conceptually, on the ability
to ``contain'' a concept scheme, along with its intra-KOS relations, as an
entity distinct from other concept schemes --- an issue, as discussed in
Section~\ref{sec:schemes} above, which the working group considered to be out of scope
for SKOS per se.  Even if a distinction between mapping relations and semantic
relations might, in principle, be anchored in a formally solid notion of
concept scheme containment, the group recognized that the evolution of concept
schemes over time could mean that related concepts in two separate schemes
could become aggregated into the same scheme, or vice versa --- situations in
which the use of formally disjoint mapping and semantic properties would prove
to be most inconvenient.  

The solution adopted by the group was to make the formal-semantic distinction
between mapping and semantic properties very weak while emphasizing the
``conventional'' difference between the two types.  As explained in the SKOS
Primer, ``By convention, mapping properties are used to represent links that
have the same intended meaning as the `standard' semantic properties, but with
a different application scope.  One might say that mapping relationships are
less \textit{inherent} to the meaning of the concepts they involve. \ldots By
convention, mapping relationships are expected to be asserted between concepts
that belong to different concept schemes.''  The authors of this paper are not
aware that the lack of a strong formal distinction between mapping and 
semantic properties has been flagged as a problem in the three years since the 
publication of SKOS as a W3C Recommendation.

The two other mapping properties, \owl{skos:closeMatch} and \owl{skos:exactMatch}, were
positioned in part as alternatives to \owl{owl:sameAs}, at the time much overused as
a mapping predicate for Linked Data.  \owl{skos:closeMatch} was intended for use with
concepts sufficiently similar to be used interchangeably in a given context.
The property was not defined as transitive in order to avoid the uncontrolled
propagation of the similarity relation to further contexts.  \owl{skos:exactMatch},
defined as a transitive sub-property of \owl{skos:closeMatch}, was intended to
express a degree of similarity close enough to justify such propagation.

\subsection{Lexical labels} 
\label{sec:labels}

\begin{framed}
Using SKOS, 
\textbf{concepts} can be {\color{gray} identified using URIs, 
}
\textbf{labeled} with lexical strings in one or more natural languages, 
assigned \textbf{notations} (lexical codes),
{\color{gray}
\textbf{documented} with various types of note, 
\textbf{linked to other concepts} and organized into informal hierarchies and association networks,
aggregated into \textbf{concept schemes}, 
grouped into labeled and/or ordered \textbf{collections}, and 
\textbf{mapped} to concepts in other schemes.
}
\end{framed}

The ability to annotate a concept for purposes of display or search is met by
properties for preferred, alternative, and hidden labels (\owl{skos:prefLabel},
\owl{skos:altLabel}, and \owl{skos:hiddenLabel}), all sub-properties of
\owl{rdfs:label}.  These properties are typically used to link an instance of
\owl{skos:Concept} to an RDF plain literal, which the working group took to
mean a character string (such as the word ``love'') combined with an optional
language tag (e.g., ``en-US'').  Note that because RDF lacked a class for RDF
Plain Literal, the property definition axiom \skosaxiom{S12} could not be expressed as a
formal range assertion in the normative RDF/XML or informative OWL 1 DL
expressions of SKOS, as shown in Table~\ref{tab:intaxioms} (see also Section~\ref{subsec:skos-owl}). 
The 2005 SKOS Core specification included properties for symbolic
labels---\owl{skos:symbol} and \owl{skos:prefSymbol}---which allowed for
resource types other than RDF plain literals as labels; however, these
were dropped for the 2009 SKOS Recommendation due to a lack of clear
requirements (see \skosissue{76} and \skosissue{180}).

The notion of ``preferred label'' derived from what the thesaurus community
calls a ``preferred term'' or ``descriptor''---i.e., a ``term specified by a
controlled vocabulary for use to represent a concept when
indexing.''\urlfoot{http://www.willpowerinfo.co.uk/glossary.htm\#preferred_term}
Preferred terms are in principle unique within the representation of a concept
scheme in a given natural language.  In order to give formal expression to this
convention, the working group operationalized the notion of language as meaning
language tag, noting that language tags can be extended to distinguish
arbitrarily specific regional variants of, say, English, Portuguese, French, or
Chinese.  The integrity condition axiom \skosaxiom{S14}, therefore, specifies that a
resource ``has no more than one value of \owl{skos:prefLabel} per language tag''
(see Section~\ref{subsec:integrity} and Table~\ref{tab:intaxioms}).  The properties for
alternative and hidden labels were intended for non-preferred indexing terms,
whether displayed to users or not.  The three labeling properties are
considered pairwise disjoint (\skosaxiom{S13}), such that assigning the same literal as
both a preferred and alternative label is formally considered an error.

In keeping with the principle of minimal ontological commitment, the SKOS
labeling properties have no explicit domain contraints. This follows the
example of Dublin Core, which does not specify domains for many of its
properties.  The lack of specific domains allows the SKOS labeling properties
to be used in contexts other than concept schemes, providing Semantic Web
applications with a generic vocabulary for labels---a usage already seen in
various OWL ontologies~\cite{manaf12:skos} and supported by non-SKOS-centric
tools such as Prot\'eg\'e.

In order to address the need for associating concepts with alphanumeric codes
such as ``M1495-2199'' (meaning ``Vocal music'' in Library of Congress
Classification), the working group introduced a property
\owl{skos:notation} (see \skosissue{79}).
A SKOS notation is intended to uniquely identify a concept within a given
concept scheme.  It differs from a lexical label ``in that a notation is not
normally recognizable as a word or sequence of words in any natural language.''
As explained in SKOS Reference, Section 6.5.1., ``By convention, the property
\owl{skos:notation} is only used with a typed literal in the object position of
the triple, where the datatype URI denotes a user-defined datatype
corresponding to a particular system of notations or classification
codes.''\urlfoot{http://www.w3.org/TR/skos-reference/\#L2613}

\paragraph{Relations between labels.} 

The ability to model binary relations between lexical labels was identified
as a candidate requirement for
SKOS (\skosrequirement{R-RelationshipsBetweenLabels}).
It should be possible, for example, to assert that the label ``FAO''
is related to the label ``Food and Agriculture Organization'' via a relation
``acronym for.''
The proposals initially considered for enabling such assertions offered
combinations of three basic ideas: creating a class for instantiating a
``term'' to which a plain-literal label could be associated;
dropping range restrictions on the SKOS labeling properties so that they could
be associated with either RDF plain literals or with instances of such a class;
and viewing relations, such as the ``acronym for'' relation, as classes.
Instances of such classes would be linked from a concept, via (for example) a
\owl{seeLabelRelation} property, and would link, via an n-ary relation
pattern~\cite{noy}, both to
a full form (ex:fullForm ``Food and Agriculture Organization'') and to an
acronym form (ex:acronymForm ``FAO''@en)---a pattern which, it was
recognized, would involve replicating the label
literals.\urlfoot{http://www.w3.org/2006/07/SWD/wiki/SkosDesign/RelationshipsBetweenLabels.html}

The solution that emerged---split off into an optional appendix, ``SKOS
eXtension for Labels (SKOS-XL),'' with its own SKOS-XL namespace URI, in order
to keep the main SKOS specification as simple as
possible\urlfoot{http://www.w3.org/TR/skos-reference/\#xl}---defined
a class, \owl{skosxl:Label}, instances of which are associated with exactly one
literal form (see SKOS-XL axiom \skosaxiom{S52} in Table~\ref{tab:xlaxioms}).  The properties
\owl{skosxl:prefLabel}, \owl{skosxl:altLabel}, and \owl{skosxl:hiddenLabel}
were coined, with the class \owl{skosxl:Label} as their range.  The property
\owl{skosxl:labelRelation} was coined as a common super-property for
applications defining their own specific label relations. The working group
felt that defining properties for specific types of label relation was out of
scope due to insufficient consensus on what would comprise a reasonably
complete set.

In order to ensure the interoperability of data created using the SKOS and
SKOS-XL labeling properties, three axioms were formulated to declare a property
chain composed of a SKOS-XL labeling property with a literal form.  For
example, the chain ``\owl{(skosxl:prefLabel, skosxl:literalForm)}'' is a
sub-property of the corresponding SKOS labeling property (in this case,
\owl{skos:prefLabel}) (see axioms \skosaxiom{S55}, \skosaxiom{S56}, and
\skosaxiom{S57} in Table~\ref{tab:xlaxioms}).  In other words, SKOS-XL labels
can be ``dumbed down'' to corresponding SKOS labels.  
It is worth noting that the \owl{skosxl:literalForm} property chain is
analogous to a pattern described in the 1999 W3C Recommendation for RDF,
whereby one of the properties of a ``structured value'' is marked, using the
property \owl{rdf:value}, as ``the principal value of the main
relation'' of a subject to a value
resource.\urlfoot{http://www.w3.org/TR/1999/REC-rdf-syntax-19990222/\#ex-NonBinary}

Defining labels, optionally, as individuals that could be annotated or related
among themselves in arbitrary ways allowed the working group to resolve an
issue raised with regard to the assertion of mapping relations between the
labels of different concept schemes (\skosissue{49}) and an issue requiring the
capability of applying annotations to the lexical items used as labels
(\skosissue{27}).  Two concerns that arose during discussions of modeling
alternatives for label relations were: identity conditions (When are two
instances of the class \owl{skosxl:Label} the same individual?), and the formal
relationship between the class \owl{skosxl:Label} and the set of RDF plain
literals (Can instances of the class \owl{skosxl:Label} have more than one
literal form?). The working group decided to assert that instances of
\owl{skosxl:Label} have exactly one literal form in order to avoid ambiguity,
but that sharing a common literal form should not be sufficient to infer that
two instances of the class \owl{skosxl:Label} were the same individual.  In
other words, two distinct instances of \owl{skosxl:Label} might have the same
literal form; there is no one-to-one mapping between the class extension of
\owl{skosxl:Label} and the set of RDF plain literals.

\subsection{Documentation Properties} \label{sec:annotations}

\begin{framed}
  Using SKOS, \textbf{concepts} can be {\color{gray} identified using URIs,
  \textbf{labeled} with lexical strings in one or more natural
  languages, assigned \textbf{notations} (lexical codes),}
  \textbf{documented} with various types of note, {\color{gray} \textbf{linked to
    other concepts} and organized into informal hierarchies and
  association networks, aggregated into \textbf{concept schemes},
  grouped into labeled and/or ordered \textbf{collections}, and
  \textbf{mapped} to concepts in other schemes.}
\end{framed}

SKOS provides a number of documentation (or note) properties. These allow for a
variety of annotations including general notes, change notes, definitions,
editorial notes, examples, historical notes, and scope notes. These seven note
types provided are not intended to be exhaustive, and it is expected that
specific application domains may extend the documentation properties
(potentially via sub-properties of the given properties, thus allowing generic
SKOS machinery access to information asserted using bespoke properties). As
with labeling properties, no domains are given for these properties, allowing
their usage outside of SKOS concept schemes. 

In addition, the documentation properties have no ranges asserted (in contrast
to labels). As discussed in the SKOS Primer,\urlfoot{http://www.w3.org/TR/skos-primer/\#secadvanceddocumentation}
this allows for a number of different documentation patterns, including the use
of literals, the use of blank nodes for structured annotations, and the use of
document references.

\subsection{Concept Collections}
\label{sec:collections}

\begin{framed}
Using SKOS, 
\textbf{concepts} can be {\color{gray} identified using URIs, 
\textbf{labeled} with lexical strings in one or more natural languages, 
assigned \textbf{notations} (lexical codes),
\textbf{documented} with various types of note, 
\textbf{linked to other concepts} and organized into informal hierarchies and association networks,
aggregated into \textbf{concept schemes}, 
}
grouped into labeled and/or ordered \textbf{collections}, and 
{\color{gray}
\textbf{mapped} to concepts in other schemes.
}
\end{framed}

In thesauri and other structured KOSs, concepts can be grouped into
semantically meaningful bundles. For example, arrays are used to group
specializations of a concept that share a common feature: the concept ``cups''
might be specialized into a first group of ``cups by form'' (``stemware'',
``tumbler''\ldots) and a second group of ``cups by function'' (``coffee cups'',
``ice cream cups''\ldots)~\cite{iso25964}. This is especially useful for
displaying KOSs: these groups are indeed most often meant as a navigation aid
in a conceptual network, not to be used for describing resources. SKOS supports
the
requirement (\skosrequirement{R-GroupingInConceptHierarchies}),
discussed in \skosissue{33} for representing
such constructs using the \owl{skos:Collection} class and its subclass
\owl{skos:OrderedCollection} for groups where the ordering of concepts
matters. 

Note that SKOS defines \owl{skos:Collection} as disjoint with
\owl{skos:ConceptScheme} and \owl{skos:Concept}. This has important
consequences. First, it can raise issues when representing ``subsets of
vocabularies'' such as micro-thesauri in the Eurovoc
thesaurus\urlfoot{http://eurovoc.europa.eu/drupal/?q=node/555} or
subdivision lists in the Library of Congress Subject
Headings\urlfoot{http://id.loc.gov/authorities/subjects}. 
The disjointness constraint forces data modelers to opt
for using (a sub-class of) either \owl{skos:Collection} or
\owl{skos:ConceptScheme}, a choice that can be hard to make in the absence
of clear guidance in the SKOS documentation. Eurovoc thus now represents
microthesauri as concept schemes, while LCSH represents subdivision lists as
collections. Fortunately, the KOS community has realized this and started to
address the problem, as witnessed by recent advocacy on how to relate ISO 25964
thesaurus standard's ``concept groups'' to \owl{skos:Collection} and
\owl{skos:ConceptScheme}~\cite{iso-skos-correspondence}. 

Note, too, that collections cannot be used in combination with semantic
relations to assign them a position in the semantic structure of a KOS.  It is
not consistent with the SKOS data model to declare
a collection to be a semantic generalization or refinement
of a ``normal'' SKOS concept with statements using \owl{skos:broader}.
In SKOS, concepts are merely grouped into collections using the properties
\owl{skos:member} and \owl{skos:memberList}. It may be seen as an
obstacle to represent simply semantic hierarchies with collections, and a
deviation from the minimal commitment approach. But it is in fact the
consequence of a conscious choice to keep data on semantic relations between
concepts clearly separate from the display-related considerations that usually motivate
the creation of collections. SKOS takes the stance that fitting collections into
KOS hierarchies must be handled by specific display algorithms that reflect the
need of users in a given navigation
environment (see \skosissue{84}).

\subsection{Issues deemed out of scope}
\label{sec:out-of-scope}

Originally chartered for just 20
months,\urlfoot{http://www.w3.org/2006/07/swdwg-charter} the Semantic
Web Deployment Working Group needed 35 months to complete its work.  In order
to focus its efforts and keep the specification as short and simple as
possible, the group declared several topics to be out of scope.

\begin{itemize}

\item{\textbf{Concept coordination}}.  Many KOSs are intended to be used as
building blocks for constructing ``coordinated'' concepts, for example to
aggregate the ``simple'' concepts ``aspirin'' and ``side effect'' into a
``compound'' concept ``aspirin -- side-effects.'' Compound concepts can be
created on a one-off basis by catalogers, as they are needed in resource
description, or they can be added as concepts to the KOS itself by its
maintainers (which is known as ``pre-coordination,'' as with the Library of Congress subject heading
``China -- history'').  The working group recognized this well-known pattern
--- ``the ability to create new concepts from existing ones, e.g. by using
special qualifiers that add a shade of meaning to a normal concept''---as a
candidate
requirement (\skosrequirement{R-ConceptCoordination}).
The group also considered a common practice in the thesaurus
world~\cite{iso2788} whereby two simple concepts (such as ``Road transport'' and
``Safety'') are designated to be used in combination instead of minting a new compound
concept such as ``Road
safety'' (see \skosissue{45}).


After much discussion, the group decided to postpone these
issues (\skosissue{40}, \skosissue{131}). While the
requirements for coordination were not questioned, the group considered them to
be relevant more for particular thesaurus and subject heading applications than
to the interchange of KOSs generally.  The group also noted that the patterns
proposed to represent concept combinations were rather complex and largely
untested. Finally, it was felt that allowing the core SKOS model to handle such
constructs could be seen as a potentially confusing move towards supporting
some functions of formal ontology languages such as OWL---languages which
support the definition of complex classes or properties from more primitive
vocabulary elements.

In retrospect, the authors feel that the decision to postpone was sound. It not
only kept untested patterns out of SKOS, avoiding delays in finalizing the
standard; it also motivated the community to tackle the issue itself.  By the
end of 2010, for example, the Library of Congress had developed a first version of
MADS/RDF~\cite{mads-rdf}, an extension to SKOS which, among other things,
supports concept coordination within library subject heading lists.

\item{\textbf{Subject indexing}}.  As defined by Leonard Will, subject indexing
involves ``intellectual analysis of the subject matter of a document to
identify the concepts represented in it, and allocation of the corresponding
preferred terms to allow the information to be
retrieved.''\urlfoot{http://www.willpowerinfo.co.uk/glossary.htm}  The
working group recognized as a candidate requirement the ``ability to represent
the indexing relationship between a resource and a concept that indexes it,''
whereby the SKOS model would include ``mechanisms to attach a given resource
(e.g. corresponding to a document) to a concept the resource is about, e.g. to
query for the resources described by a given
concept'' (see \skosrequirement{R-IndexingRelationship}, \skosissue{77}).
Noting the existence of indexing relation properties in other vocabularies,
such as Dublin Core's \owl{dc:subject}, the working group declared such
properties to be out of scope and decided not to carry forward the 
property \owl{skos:subject} from the 2005 SKOS
Core specification.\urlfoot{http://www.w3.org/TR/2005/WD-swbp-skos-core-spec-20051102/\#subject}
(It should be noted that recent Web crawls show
\owl{skos:subject} to be one of the most-used properties from SKOS, due
primarily to its use in
DBPedia.\urlfoot{http://sindice.com/search?q=&nq=\%28*\%20\%3Chttp\%3A\%2F\%2Fpurl.org\%2Fdc\%2Fterms\%2Ftitle\%3E\%20*\%29&fq=&interface=advanced})
For lack of a SKOS
indexing vocabulary, a candidate requirement for distinguishing between
indexing and non-indexing concepts was also declared out of
scope (see \skosrequirement{R-IndexingAndNonIndexingConcepts}, \skosissue{46}).

\item{\textbf{Provenance information about mappings}}.  The ability ``to record
provenance information on mappings between concepts in different concept
schemes'' was recognized as a candidate requirement for
SKOS (\skosrequirement{R-MappingProvenanceInformation}).
The issue was resolved with a decision not to introduce specific SKOS
vocabulary about the provenance of mappings (\skosissue{47}).  Rather, the group felt that this
issue depended on the use of standard containment mechanisms for encompassing
mapping assertions within a context that could be denoted with a URI---an
issue relevant for RDF in general, specifically for the future development of
standards regarding ``named graphs'' (see also the discussion of containment in
Section ~\ref{sec:schemes}).

\item{\textbf{Describing concept schemes}}.  Concept schemes have authors,
titles, publishers, dates issued, subject coverage, and the like.  The working
group felt that the question of what properties to use in describing a concept
scheme was an issue best left to communities of practice.  Shortly after the
publication of SKOS in 2009, for example, a joint DCMI--NKOS task group was
formed between the Dublin Core Metadata Initiative and Networked Knowledge
Organization Systems community to develop an application profile and a KOS Type
Vocabulary for describing
KOSs.\urlfoot{http://dublincore.org/groups/nkos/}

\item{\textbf{Concept evolution}}.  The working group acknowledged the
importance of mechanisms for representing the temporal evolution of concept
schemes---an issue that raises questions of granularity (whether to version
individual statements, concept descriptions, or entire concept schemes) and of
how to represent such versioning information in interoperably machine-readable
ways.  The group considered this topic best left to the community for research
and testing (see~\cite{versioning}).

\end{itemize}


%% file: semantics.tex
\section{Formal semantics}
\label{sec:sem}

This section discusses aspects of SKOS relating to its formal semantics, in
particular highlighting the use of OWL. 
The working group\footnote{In this section, references to ``the working
  group'' refer to the Semantic Web Deployment Working Group. Other
  working groups will be referred to by their full name.} was tasked to specify SKOS in accordance with OWL,
so as to allow for applications to validate SKOS datasets or to
infer new facts from the ones explicitly encoded by publishers of SKOS data.
The SKOS model is thus specified by defining OWL classes and properties, which
can be interpreted using OWL's formal semantics. A particular SKOS concept
scheme is an instantiation of the OWL ontology that defines SKOS in which SKOS
concepts are instances of the class \owl{skos:Concept} with characteristics expressed
using the SKOS properties.

\subsection{Axioms supporting inference}
\label{subsec:inferenceaxioms}

As described above, the SKOS data model contains a number of
axioms (stated as \skosaxiom{S1} to \skosaxiom{S46} in the Recommendation\footnote{SKOS-XL
includes additional axioms \skosaxiom{S47} to \skosaxiom{S62}}) relating to the classes and
properties of the SKOS vocabulary.

All but six of these axioms, as listed in Table~\ref{tab:defaxioms}, describe how the classes and
properties of SKOS are defined, primarily by stating subclass or sub-property
relationships or domain and range assertions. These axioms allow the use of
inference engines (``reasoners'') to derive additional information about the
nature of, and relationships among, components of a concept
scheme. Note that such inference concerns the concept scheme as an information
artefact in itself and says nothing about the nature of the resources or
``real-world'' entities to which the concepts of a concept scheme may refer.
For example, the axiom \skosaxiom{S4} allows the inference that an object of a triple using
\owl{skos:inScheme} is an instance of the
class \owl{skos:ConceptScheme}.\footnote{Note the semantics of
  \owl{rdf:range} here. A common misconception is that a concept scheme
  that does not explicitly type an object of an \owl{skos:inScheme} as
a \owl{skos:ConceptScheme} would be in error. This is not the case though---\owl{rdfs:range} 
assertions are not \emph{constraints}, but are
conditions on interpretations providing inferences.}
Axiom \skosaxiom{S25} allows an application that is OWL-semantics aware to infer the
presence of \owl{skos:broader} relationships in a concept scheme that asserts only \owl{skos:narrower} relationships. 

The SKOS-XL extension (see Section\ref{sec:labels}) includes axioms relating to property chains, for example \skosaxiom{S55} would allow an application given the triples:

\begin{quote}
\begin{verbatim}
ex:concept-1234 skosxl:prefLabel ex:label-5678.
ex:label-5678 skosxl:literalForm "love".
\end{verbatim}
\end{quote}

to infer the triple
\begin{quote}
\begin{verbatim}
ex:concept-1234 skos:prefLabel "love".
\end{verbatim}
\end{quote}

\subsection{Integrity conditions} \label{subsec:integrity}

In addition to the axioms described above, a number of \emph{integrity conditions}
(labeled as \skosaxiom{S9}, \skosaxiom{S13}, \skosaxiom{S14}, \skosaxiom{S27}, \skosaxiom{S37}, and \skosaxiom{S46}) are also given. The
integrity conditions serve a different purpose to the other axioms
stated, in that they are intended to facilitate and promote
interoperability by defining circumstances under which data are not
consistent with respect to the SKOS data model. 
Details of integrity conditions are given in Table~\ref{tab:intaxioms}.

The working group was chartered to create a machine-readable specification of the SKOS axioms using the OWL language, which forms the base for exchanging and exploiting formal specifications of ontologies on the Web of Data, as envisioned in the W3C Semantic Web technology stack.
The SKOS Recommendation makes no assumptions, however, as to
\emph{how} implementation of the checking of integrity conditions for
a particular concept scheme are performed. They could be checked
through inference, but other mechanisms could be used, for example
querying for particular graph patterns or the use of rule driven approaches such as SPIN\urlfoot{http://spinrdf.org/} or Pellet's Integrity Constraints.\urlfoot{http://clarkparsia.com/pellet/icv}

\subsection{SKOS as an OWL Ontology} \label{subsec:skos-owl}

\paragraph{Historical context.}

The Web Ontology Language (OWL) was first published as a collection of W3C
Recommendations in 2004\urlfoot{http://www.w3.org/2004/OWL/} developed by the
Web Ontology working group,\urlfoot{http://www.w3.org/2001/sw/WebOnt/} first
convened in 2001. One key aspect of OWL was the definition of three
sublanguages known as OWL Lite, OWL DL and OWL Full. OWL DL supported those
users who wanted maximum expressiveness while still retaining computational
completeness. OWL Full provided greater expressiveness and syntactic freedom,
but with a lack of computational guarantees. OWL Lite was a subset of OWL DL
intended to support users needing a classification hiearchy and simple
constraints. The working group was tasked to specify SKOS in accordance
with OWL Full.

In 2007, the OWL Working Group\urlfoot{http://www.w3.org/2007/OWL}
was convened, with a charter to produce an update to OWL, resulting
(in 2009) in a collection of recommendations defining
OWL 2\urlfoot{http://www.w3.org/TR/owl2-overview/} (also earlier known
as OWL 1.1 during the process). The work of the OWL Working Group
overlapped with the work of the Semantic Web Deployment Group, with
the consequence that the SKOS recommendation did not have the
opportunity of using OWL 2 features in the SKOS recommendation (this
point and the related issue of defining
SKOS within the limits of OWL DL in is covered in more detail in
Section~\ref{subsec:owl-dl-compatibility}). 

To avoid confusion, this section refers explicitly
to the original (2004) recommendation as OWL 1 and the revision (2009)
as OWL 2. 

\paragraph{SKOS as an OWL ontology.}
\label{subsec:skosasowlontology}

The SKOS data model is represented as an OWL 1 ontology, i.e. a collection of classes and properties with associated axioms. 

The SKOS Namespace Document RDF/XML
Variant\urlfoot{{http://www.w3.org/TR/skos-reference/skos.rdf}}
provides definitions of the classes and properties of this model
using OWL 1, along with axioms that represent integrity conditions on
the data represented using SKOS. As there
are limits to the expressivity of OWL 1 (and of its subspecies or
fragments), not all of the desired constraints can be fully expressed
using OWL 1. This is further discussed below. Where this is the
case, the constraint is expressed as a comment in the schema.

\subsection{Compatibility with OWL 1 DL and OWL 2}
\label{subsec:owl-dl-compatibility}

SKOS is defined as an OWL 1 ontology, and a
requirement \skosrequirement{R-CompatibilityWithOWL-DL}
made to the working group was that SKOS should provide a legal OWL 1 DL
ontology, primarily to ensure compatibility with editing tools and to
facilitate the use of reasoners, many of which operate in the OWL 1 DL
space.

This was problematic as OWL 1 DL lacked the expressivity needed to
capture some of the assertions. For example, OWL 1 DL has no facility
to express hierarchies of annotation properties. Nor does OWL 1
provide a mechanism for stating axioms concerning property chains as
used in axiom \skosaxiom{S55}. A further complication was that the
work of the Semantic Web Deployment Group overlapped with that of the
OWL Working Group, which was defining the OWL 2 Recommendation (also
referred to as OWL 1.1 during the process). OWL 2 was likely to
introduce features that would support some of these assertions, but as
the OWL Working Group was scheduled to finish \emph{after} SKOS
delivery, the normative SKOS reference could not make reference to OWL
2. For example, the particular feature supporting hierarchies of
annotation properties was ultimately introduced into OWL 2.

In order to provide some support for reasoning engines and those applications
working in the OWL 1 DL space, a
``pruned'' RDF schema was produced, providing a \textbf{non-normative}
resource. This is made available (in a non-normative fashion) as the SKOS RDF
Schema - OWL 1 DL
Sub-set.\urlfoot{http://www.w3.org/TR/skos-reference/skos-owl1-dl.rdf} In
particular, the pruning removed axioms stating that SKOS labeling properties
are sub-properties of \owl{rdfs:label} as sub-property axioms are not applicable
to annotation properties in OWL 1 DL. 

This particular pruning of the schema in order to provide a
valid OWL 1 DL ontology is only one of a number of possible ways in
which the OWL 1 Full RDF Schema for SKOS can be adjusted in order to sit
in the OWL 1 DL space---each of which would have differing semantic
consequences. As a result the OWL 1 DL prune was considered
non-normative.

Other constraints were also problematic in terms of OWL 1
representation. \skosaxiom{S14} states that ``A resource has no more than one
value of \owl{skos:prefLabel} per language tag.''
This was not expressible in OWL 1. Nor were property
disjointness constraints as expressed in \skosaxiom{S13}, \skosaxiom{S27} and \skosaxiom{S46}. Issues
relating to compatibility with OWL 1 DL---\skosissue{38}, \skosissue{137}, \skosissue{138}---were thus formally postponed by
the working group, indicating that, should work resume on an updated
recommendation, this should be the focus of attention. 

Comments from members of the OWL Working Group (raised as
\skosissue{155} and \skosissue{157}) highlighted areas where an
adjustment to the model would potentially provide better alignment
with the emerging OWL 2 recommendation. \skosissue{157} was formally
postponed. Following the resolution of \skosissue{135}, labeling properties were defined as \owl{owl:AnnotationProperty}.

\subsection{Machine-readable Formalizations, Formal Semantics and Data Quality}
\label{subsec:machine-formalised}

As discussed earlier, the SKOS data model is represented as a
collection of axioms, some providing definitions of classes and
properties, which then support inference, others asserting integrity
conditions. When representing these axioms in a machine-readable way
(which was the main mission of the SWD group), the implementation
creates two ``layers'' orthogonal to this question of definition versus
integrity:

\begin{enumerate}
\item Axioms formally represented in the ontology, for example, sub-property relations to \owl{skos:semanticRelation};
\item Axioms that are not explicitly represented in the ontology, primarily due to a lack of expressivity in the representations, for example assertions about disjoint properties.
\end{enumerate}

The SKOS RDF/OWL representation thus proposes a core layer for inference and
validation of SKOS data.  However, as mentioned in
Section~\ref{subsec:integrity}, the working group did not
assume a specific technique for checking the integrity conditions of the SKOS
data model.  

This flexibility can be explained by the difficulty of representing all
integrity conditions in the OWL language (as discussed above). But it is also
in line with a more fundamental stance of the SWD group, which allows for a
flexible approach to data quality in SKOS, generally. In addition to the two
layers described (formal versus informal axioms),
the SKOS reference includes what one might call \emph{guidelines} which are
weaker recommendations, for example that \owl{skos:closeMatch} should be used
to relate concepts from different schemes. There is no attempt at formal
representation of the latter, nor is it considered an integrity constraint that
should not be violated. These assertions are more
``advisory,'' but are still somehow part of the SKOS model.  It is
left to
SKOS implementations to adapt these guidelines---or others from specific
domains, such as thesaurus design guidelines~\cite{iso2788}, which can provide
useful ``checks'' for SKOS data.

Example approaches to validation of SKOS data include the Poolparty Thesaurus Consistency Checker,\urlfoot{http://poolparty.punkt.at/} which runs custom validation rules derived from the SKOS axioms. The qSKOS tool by Mader et al.~\cite{mader12:quality} is used to identify a number of quality issues in SKOS vocabularies, in
particular fifteen ``guideline'' violations. The Skosify
tool~\cite{suominen12:skosify} identifies an overlapping (but slightly
different) set of criteria, some of which correspond to SKOS integrity
conditions (e.g., \skosaxiom{S13} concerning disjointness of alternate
and preferred labels). 

Tables~\ref{tab:defaxioms}, \ref{tab:intaxioms} and \ref{tab:xlaxioms}
provide a summary of the axioms in the SKOS and SKOS-XL data
models. It also highlights those axioms that lack a formal machine
representation in either the normative RDF Schema or the non-normative
OWL 1 DL prune (note there is no corresponding OWL 1 DL prune of SKOS-XL). 

\input{axiomtable}

\subsection{SKOS Namespace URI}

A question that was the focus of much attention during the
Recommendation process was that of the URI to be used for SKOS, formally raised as \skosissue{153} and \skosissue{175}. Earlier
work from the Semantic Web Best Practices and Deployment Group
resulted in a SKOS Core Working
Draft.\urlfoot{http://www.w3.org/TR/swbp-skos-core-spec} This
Working Draft was a key input to the work of the working group, and much of the content of the original
Core was preserved in the final Recommendation. The original core
defined vocabulary using the namespace URI
\url{http://www.w3.org/2004/02/skos/core}. 

Various possibilities were open to the working group:

\begin{enumerate}
\item Provide a new namespace URI for the SKOS Recommendation;
\item \label{namespace:choice} Use the existing SKOS Core namespace URI for the SKOS Recommendation,
  potentially redefining or changing the semantics of URIs defined in that
  namespace.
\item Use the existing SKOS Core namespace URI for the SKOS Recommendation,
  minting new URIs for those vocabulary elements where semantics had
  been changed. 
\end{enumerate}

As an example of a situation where the semantics of a vocabulary
element had changed, consider the hierarchical semantic relations
\owl{skos:broader} and \owl{skos:narrower}. 

In the original core, these properties
were declared to be transitive, while in the final SKOS
recommendation, as discussed in Section~\ref{sec:transitivity}, they
were not (instead a transitive reduction~\cite{neon:transitive} design
pattern was used, introducing transitive superproperties
\owl{skos:broaderTransitive} and \owl{skos:narrowerTransitive}).

Each option had pros and cons. The introduction of a new namespace URI would reduce
the problems of inconsistent interpretations of existing vocabularies that may
have been producing using the original semantics. However, a new namespace URI
would then potentially require changes to existing tools,
infrastructure and
concept schemes.

The final decision made was for option~\ref{namespace:choice}. It was
felt that disruption to the existing body of data that had been
published using SKOS Core would have been significant if the
namespace URI or property names of key elements have been
changed. Although this resulted in a change of semantics to some
properties, applications should, in principle, be able to make use of
the machine-readable published schema to access those semantics. 

On a similar note, elements were removed from the SKOS Core vocabulary (see
discussion in the SKOS
Reference\urlfoot{http://www.w3.org/TR/skos-reference/\#namespace}) although
historical versions of the schemas remain
available.\urlfoot{http://www.w3.org/2004/02/skos/history}
 
Although SKOS Core had at that point been deployed by early adopters
for several years, changing the semantics associated with the URIs was
unproblematic strictly from the standpoint of process because the 2005
specification had only attained the status of Working Draft---a type
of specification by definition subject to change.


It is worth noting that following publication of SKOS as a
Recommendation, it was observed that the the SKOS OWL 1 DL prune
ontology had no version IRI, thus breaking a rule specified in the
2009 OWL 2 recommendation, that ``If an ontology has an ontology IRI
but no version IRI, then a different ontology with the same ontology
IRI but no version IRI SHOULD NOT exist.'' In order to address this,
an additional \owl{owl:versionIRI} triple was added to the
ontology\urlfoot{http://www.w3.org/2006/07/SWD/SKOS/reference/20090811-errata}.



%% file: axiomtable.tex
\newcommand{\absent}{\cellcolor{red}}
\newcommand{\present}{\cellcolor{green}}
\newcommand{\tablekey}{A green cell denotes that an axiom is present in
 the corresponding formalisation, while a red cell denotes absence.}

\begin{table}[ht]
\scriptsize
  \centering
  \begin{tabular}{|l|p{0.8\textwidth}|c|c|}
\hline
\rotatebox{75}{\textbf{Axiom}} & \rotatebox{75}{\textbf{Content}} & \rotatebox{75}{\textbf{RDF Schema}} & \rotatebox{75}{\textbf{OWL Prune}} \\
\hline
\skosaxiom{S1} & \raggedright     \owl{skos:Concept} is an instance of \owl{owl:Class}. & \present & \present \\
\hline
\skosaxiom{S2} & \raggedright     \owl{skos:ConceptScheme} is an instance of \owl{owl:Class}. & \present & \present \\
\hline
\skosaxiom{S3} & \raggedright     \owl{skos:inScheme}, \owl{skos:hasTopConcept} and \owl{skos:topConceptOf} are each instances of \owl{owl:ObjectProperty}. & \present & \present \\
\hline
\skosaxiom{S4} & \raggedright     The \owl{rdfs:range} of \owl{skos:inScheme} is the class \owl{skos:ConceptScheme}. & \present & \present \\
\hline
\skosaxiom{S5} & \raggedright     The \owl{rdfs:domain} of \owl{skos:hasTopConcept} is the class \owl{skos:ConceptScheme}. & \present & \present \\
\hline
\skosaxiom{S6} & \raggedright     The \owl{rdfs:range} of \owl{skos:hasTopConcept} is the class \owl{skos:Concept}. & \present & \present \\
\hline
\skosaxiom{S7} & \raggedright     \owl{skos:topConceptOf} is a sub-property of \owl{skos:inScheme}. & \present & \present \\
\hline
\skosaxiom{S8} & \raggedright     \owl{skos:topConceptOf} is \owl{owl:inverseOf} the property \owl{skos:hasTopConcept}. & \present & \present \\
\hline
\skosaxiom{S10} & \raggedright    \owl{skos:prefLabel}, \owl{skos:altLabel} and \owl{skos:hiddenLabel} are each instances of \owl{owl:AnnotationProperty}. & \present & \present \\
\hline
\skosaxiom{S11} & \raggedright    \owl{skos:prefLabel}, \owl{skos:altLabel} and \owl{skos:hiddenLabel} are each sub-properties of \owl{rdfs:label}. & \present & \absent\\
\hline
\skosaxiom{S12} & \raggedright    The \owl{rdfs:range} of each of \owl{skos:prefLabel}, \owl{skos:altLabel} and \owl{skos:hiddenLabel} is the class of RDF plain literals. & \absent& \absent \\
\hline
\skosaxiom{S15} & \raggedright    \owl{skos:notation} is an instance of \owl{owl:DatatypeProperty}. & \present & \present \\
\hline
\skosaxiom{S16} & \raggedright    \owl{}\owl{skos:note}, \owl{skos:changeNote}, \owl{skos:definition}, \owl{skos:editorialNote}, \owl{skos:example}, \owl{skos:historyNote} and \owl{skos:scopeNote} are each instances of \owl{owl:AnnotationProperty}. & \present & \present \\
\hline 
\skosaxiom{S17} & \raggedright    \owl{skos:changeNote}, \owl{skos:definition}, \owl{skos:editorialNote}, \owl{skos:example}, \owl{skos:historyNote} and \owl{skos:scopeNote} are each sub-properties of \owl{skos:note}.
& \present & \absent \\
\hline 
\skosaxiom{S18} & \raggedright    \owl{skos:semanticRelation}, \owl{skos:broader}, \owl{skos:narrower}, \owl{skos:related}, \owl{skos:broaderTransitive} and \owl{skos:narrowerTransitive} are each instances of \owl{owl:ObjectProperty}.
& \present & \present \\
\hline 
\skosaxiom{S19} & \raggedright    The \owl{rdfs:domain} of \owl{skos:semanticRelation} is the class \owl{skos:Concept}. & \present & \present \\
\hline
\skosaxiom{S20} & \raggedright    The \owl{rdfs:range} of \owl{skos:semanticRelation} is the class \owl{skos:Concept}. & \present & \present \\
\hline
\skosaxiom{S21} & \raggedright    \owl{skos:broaderTransitive}, \owl{skos:narrowerTransitive} and \owl{skos:related} are each sub-properties of \owl{skos:semanticRelation}. & \present & \present \\
\hline
\skosaxiom{S22} & \raggedright    \owl{skos:broader} is a sub-property of \owl{skos:broaderTransitive}, and \owl{skos:narrower} is a sub-property of \owl{skos:narrowerTransitive}. & \present & \present \\
\hline
\skosaxiom{S23} & \raggedright    \owl{skos:related} is an instance of \owl{owl:SymmetricProperty}. & \present & \present \\
\hline
\skosaxiom{S24} & \raggedright    \owl{skos:broaderTransitive} and \owl{skos:narrowerTransitive} are each instances of \owl{owl:TransitiveProperty}. & \present & \present \\
\hline
\skosaxiom{S25} & \raggedright    \owl{skos:narrower} is \owl{owl:inverseOf} the property \owl{skos:broader}. & \present & \present \\
\hline
\skosaxiom{S26} & \raggedright    \owl{skos:narrowerTransitive} is \owl{owl:inverseOf} the property \owl{skos:broaderTransitive}. & \present & \present \\
\hline
\skosaxiom{S28} & \raggedright    \owl{skos:Collection} and \owl{skos:OrderedCollection} are each instances of \owl{owl:Class}. & \present & \present \\
\hline
\skosaxiom{S29} & \raggedright    \owl{skos:OrderedCollection} is a sub-class of \owl{skos:Collection}. & \present & \present \\
\hline
\skosaxiom{S30} & \raggedright    \owl{skos:member} and \owl{skos:memberList} are each instances of \owl{owl:ObjectProperty}. & \present & \present \\
\hline
\skosaxiom{S31} & \raggedright    The \owl{rdfs:domain} of \owl{skos:member} is the class \owl{skos:Collection}. & \present & \present \\
\hline
\skosaxiom{S32} & \raggedright    The \owl{rdfs:range} of \owl{skos:member} is the union of classes \owl{skos:Concept} and \owl{skos:Collection}. & \present & \present \\
\hline
\skosaxiom{S33} & \raggedright    The \owl{rdfs:domain} of \owl{skos:memberList} is the class \owl{skos:OrderedCollection}. & \present & \present \\
\hline
\skosaxiom{S34} & \raggedright    The \owl{rdfs:range} of \owl{skos:memberList} is the class \owl{rdf:List}. & \present & \absent \\
\hline
\skosaxiom{S35} & \raggedright    \owl{skos:memberList} is an instance of \owl{owl:FunctionalProperty}. & \present & \present \\
\hline
\skosaxiom{S36} & \raggedright    For any resource, every item in the list given as the value of the \owl{skos:memberList} property is also a value of the \owl{skos:member} property. & \absent & \absent \\
\hline
\skosaxiom{S38} & \raggedright    \owl{skos:mappingRelation}, \owl{skos:closeMatch}, \owl{skos:exactMatch}, \owl{skos:broadMatch}, \owl{skos:narrowMatch} and \owl{skos:relatedMatch} are each instances of \owl{owl:ObjectProperty}.
& \present & \present \\
\hline
\skosaxiom{S39} & \raggedright    \owl{skos:mappingRelation} is a sub-property of \owl{skos:semanticRelation}. & \present & \present \\
\hline
\skosaxiom{S40} & \raggedright    \owl{skos:closeMatch}, \owl{skos:broadMatch}, \owl{skos:narrowMatch} and \owl{skos:relatedMatch} are each sub-properties of \owl{skos:mappingRelation}. & \present & \present \\
\hline
\skosaxiom{S41} & \raggedright    \owl{skos:broadMatch} is a sub-property of \owl{skos:broader}, \owl{skos:narrowMatch} is a sub-property of \owl{skos:narrower}, and \owl{skos:relatedMatch} is a sub-property of \owl{skos:related}.
& \present & \present \\
\hline 
\skosaxiom{S42} & \raggedright    \owl{skos:exactMatch} is a sub-property of \owl{skos:closeMatch}. & \present & \present \\
\hline
\skosaxiom{S43} & \raggedright    \owl{skos:narrowMatch} is \owl{owl:inverseOf} the property \owl{skos:broadMatch}. & \present & \present \\
\hline
\skosaxiom{S44} & \raggedright    \owl{skos:relatedMatch}, \owl{skos:closeMatch} and \owl{skos:exactMatch} are each instances of \owl{owl:SymmetricProperty}. & \present & \present \\
\hline
\skosaxiom{S45} & \raggedright    \owl{skos:exactMatch} is an instance of \owl{owl:TransitiveProperty}. & \present & \present \\
\hline
 \end{tabular}
 \caption{SKOS Class and Property Definition Axioms. \tablekey}
 \label{tab:defaxioms}
\end{table}

\begin{table}[ht]
\scriptsize
  \centering
  \begin{tabular}{|l|p{0.8\textwidth}|c|c|}
\hline
\rotatebox{75}{\textbf{Axiom}} & \rotatebox{75}{\textbf{Content}} & \rotatebox{75}{\textbf{RDF Schema}} & \rotatebox{75}{\textbf{OWL Prune}} \\
\hline
\skosaxiom{S9} & \raggedright     \owl{skos:ConceptScheme} is disjoint with \owl{skos:Concept}. & \present & \present \\
\hline
\skosaxiom{S13} & \raggedright    \owl{skos:prefLabel}, \owl{skos:altLabel} and \owl{skos:hiddenLabel} are pairwise disjoint properties. & \absent & \absent \\
\hline
\skosaxiom{S14} & \raggedright    A resource has no more than one value of \owl{skos:prefLabel} per language tag. & \absent & \absent \\
\hline
\skosaxiom{S27} & \raggedright    \owl{skos:related} is disjoint with the property \owl{skos:broaderTransitive}. & \absent & \absent \\
\hline
\skosaxiom{S37} & \raggedright    \owl{skos:Collection} is disjoint with each of \owl{skos:Concept} and \owl{skos:ConceptScheme}. & \present & \present \\
\hline
\skosaxiom{S46} & \raggedright    \owl{skos:exactMatch} is disjoint with each of the properties \owl{skos:broadMatch} and \owl{skos:relatedMatch}. & \absent & \absent \\
\hline
 \end{tabular}
  \caption{SKOS Integrity Condition Axioms. \tablekey}
  \label{tab:intaxioms}
\end{table}

\begin{table}[ht]
\scriptsize
  \centering
  \begin{tabular}{|l|p{0.8\textwidth}|l|l|}
\hline
\rotatebox{75}{\textbf{Axiom}} & \rotatebox{75}{\textbf{Content}} & \rotatebox{75}{\textbf{RDF Schema}} \\
\hline

\hline
\skosaxiom{S47} & \raggedright	\owl{skosxl:Label} is an instance of \owl{owl:Class}. & \present \\
\hline
\skosaxiom{S48} & \raggedright	\owl{skosxl:Label} is disjoint with each of \owl{skos:Concept}, \owl{skos:ConceptScheme} and \owl{skos:Collection}. & \present \\
\hline
\skosaxiom{S49} & \raggedright	\owl{skosxl:literalForm} is an instance of \owl{owl:DatatypeProperty}. & \present \\
\hline
\skosaxiom{S50} & \raggedright	The \owl{rdfs:domain} of \owl{skosxl:literalForm} is the class \owl{skosxl:Label}. & \present \\
\hline
\skosaxiom{S51} & \raggedright	The \owl{rdfs:range} of \owl{skosxl:literalForm} is the class of RDF plain literals. & \present \\
\hline
\skosaxiom{S52} & \raggedright	\owl{skosxl:Label} is a sub-class of a restriction on \owl{skosxl:literalForm} cardinality exactly 1. & \present \\
\hline
\skosaxiom{S53} & \raggedright	\owl{skosxl:prefLabel}, \owl{skosxl:altLabel} and \owl{skosxl:hiddenLabel} are each instances of \owl{owl:ObjectProperty}. & \present \\
\hline
\skosaxiom{S54} & \raggedright	The \owl{rdfs:range} of each of \owl{skosxl:prefLabel}, \owl{skosxl:altLabel} and \owl{skosxl:hiddenLabel} is the class \owl{skosxl:Label}. & \present \\
\hline
\skosaxiom{S55} & \raggedright	The property chain (\owl{skosxl:prefLabel}, \owl{skosxl:literalForm}) is a sub-property of \owl{skos:prefLabel}. & \absent \\
\hline
\skosaxiom{S56} & \raggedright	The property chain (\owl{skosxl:altLabel}, \owl{skosxl:literalForm}) is a sub-property of \owl{skos:altLabel}. & \absent \\
\hline
\skosaxiom{S57} & \raggedright	The property chain (\owl{skosxl:hiddenLabel}, \owl{skosxl:literalForm}) is a sub-property of \owl{skos:hiddenLabel}. & \absent \\
\hline
\skosaxiom{S58} & \raggedright	\owl{skosxl:prefLabel}, \owl{skosxl:altLabel} and \owl{skosxl:hiddenLabel} are pairwise disjoint properties. & \present \\
\hline
\skosaxiom{S59} & \raggedright	\owl{skosxl:labelRelation} is an instance of \owl{owl:ObjectProperty}. & \present \\
\hline
\skosaxiom{S60} & \raggedright	The \owl{rdfs:domain} of \owl{skosxl:labelRelation} is the class \owl{skosxl:Label}. & \present \\
\hline
\skosaxiom{S61} & \raggedright	The \owl{rdfs:range} of \owl{skosxl:labelRelation} is the class \owl{skosxl:Label}. & \present \\
\hline
\skosaxiom{S62} & \raggedright	\owl{skosxl:labelRelation} is an instance of \owl{owl:SymmetricProperty}. & \present \\
\hline

  \end{tabular}
  \caption{SKOS XL Axioms. \tablekey}
  \label{tab:xlaxioms}
\end{table}

%% file: conclusion.tex
\section{Conclusion}
\label{sec:concl}

The intellectual roots of Knowledge Organization Systems go back decades, even
centuries.  The goal of expressing Knowledge Organization Systems in a
generically interoperable way was raised already as a goal when W3C working
groups began developing the Semantic Web language, Resource Description
Framework (RDF), in the late 1990s.  In the twelve years from the beginnings of
RDF in 1997 through the finalization of SKOS as a W3C Recommendation in 2009,
the torch for this work was passed among a succession of UK and European
research projects and of W3C working groups, each of which added features,
dropped others, and progressively clarified its underlying concepts.  This
process illustrated the challenge of developing specifications that depend on
related specifications which, in today's continually evolving environment, are
inevitably subject to change. As discussed above, SKOS would have looked
slightly different if OWL 2, published as a W3C Recommendation just two months
after SKOS,\urlfoot{http://www.w3.org/TR/owl2-overview/} had been
finalized just half a year earlier.

This paper highlights a number of issues that were ``postponed'' -- a status
which marks them as being of potential interest to future working groups:

\begin{itemize}

\item{\textbf{SKOS and OWL.}} In the three years since the publication of the W3C Recommendation for SKOS,
one of the most Frequently Asked Questions has been that of the relationship
between information KOSs, expressed using SKOS, and OWL ontologies.  As
discussed in Section~\ref{sec:concepts}, almost anything can be considered a SKOS
Concept (as long as it is not a SKOS Concept Scheme, Collection, or Label), and
\owl{foaf:focus} provides a way to link SKOS concepts to things in the world to which
those concepts refer.  The working group defined SKOS this way so as not to
preclude experimentation with usage patterns as yet unforeseen.

\item{\textbf{The formal expression of SKOS.}}  The formal expression of SKOS
axioms could be enhanced in light of OWL 2 (see \skosissue{38},
\skosissue{136}, \skosissue{137}, \skosissue{138}, \skosissue{155}, and
Section~\ref{subsec:owl-dl-compatibility}).  Such enhancements could help
consolidate a more consistent approach to validating SKOS concept schemes.  The
choice of axioms pruned to create the non-normative OWL DL 1 Prune should also
be revisited in light of implementation experience (see Section
\ref{subsec:owl-dl-compatibility}).  A future working group might also want to
formulate recommendations on the use of non-OWL semantics based, perhaps, on
constraints with a closed-world interpretation.

\item{\textbf{Inference, validation, and quality control.}}
A future working group might want to incorporate work being done in the
implementation community on extending the definition of ``conformance,'' for
example to enhance support for specific types of KOS (see \skosissue{35}).  

\item{\textbf{Extending SKOS with additional properties.}}  Extending SKOS with
richer semantics relations (\skosissue{56}, \skosissue{149}, \skosissue{150},
and \skosissue{178}) remains a very popular topic and has resulted already in
proposals, e.g., in the ISO 25964 standard~\cite{iso25964}.  The issue of
extending SKOS with symbolic labels remains unaddressed, though the authors see
the potential for experimentation related, for example, to Web accessibility.
The potential refinement of SKOS mapping properties perhaps awaits a stronger
push on formulating best practice for mapping in the Semantic Web context
generally (see \skosissue{176}).  

\item{\textbf{Concept coordination.}}  Patterns defined by MADS and ISO 25964
for concept coordination could be evaluated in light of implementation
experience, especially as pre- and post-coordination patterns are tested in the
context of different types of KOS and in information retrieval applications
(see \skosissue{40}, \skosissue{45}, \skosissue{131}, and
Section~\ref{sec:out-of-scope}).

\item{\textbf{Concept scheme containment and provenance.}}  As of late 2012,
the RDF Working Group is working towards standardizing an approach to naming
graphs and datasets.  As pointed out in Section \ref{sec:schemes}, the
identification of graphs is relevant to all issues which require that concept
schemes be delimited, or ``contained,'' for the purpose of tracking provenance
or expressing precise alignments (see
\skosrequirement{R-ConceptSchemeContainment} and
\skosrequirement{R-MappingProvenanceInformation}).  The Linked Data community
is developing relevant practices and vocabularies, such as
VoID\urlfoot{http://vocab.deri.ie/void}, which addresses the provenance both of
generic datasets and of more specific ``linksets.'' The SKOS community has made
some progress in the past years on modeling concept
evolution\urlfoot{http://www.w3.org/2001/sw/wiki/SKOS/Issues/ConceptEvolution}---an issue of particular interest to builders of SKOS
``registries'' and APIs.

\item{\textbf{Best practices for modeling SKOS Concept Schemes.}}  As discussed
in Section~\ref{sec:collections} there are some concept groupings, such as
micro-thesauri, to which the architecture of SKOS Collections versus SKOS
Concept Schemes does not neatly fit, suggesting a need to clarify best
practices.

\end{itemize}

The W3C Semantic Web Deployment Working Group, which carried SKOS forward
during the final three years of this process, began its work with a draft
specification, at the time called SKOS Core, which had already been widely
deployed and tested by early adopters.  In the three years since its
publication in 2009, SKOS has become one of the most widely used vocabularies
in the Linked Data
cloud\urlfoot{http://thedatahub.org/dataset?tags=format-skos}---a
context to which its flexible, generic design, based on the principle of
minimal ontological commitment, is uniquely well-suited.  As its designers
intended, SKOS continues to be adapted and extended to meet more specialized
requirements.  Some of this work is discussed on the public-esw-thes mailing list\urlfoot{http://lists.w3.org/Archives/Public/public-esw-thes}
and tracked on the W3C SKOS community wiki\urlfoot{http://www.w3.org/2001/sw/wiki/SKOS}.
The development of SKOS has been the collective result of several
dozen contributors working, typically, in the context of working groups or
projects of two or three years' duration.  The accumulated impact of such
incremental contributions becomes clear only in retrospect, looking back with
the perspective of a decade or two.


%% file: acks.tex
\subsection*{Acknowledgements}

The authors would like to acknowledge the contributions of all of the members of the W3C Semantic Web Deployment Working Group in the production of SKOS. 